\pdfoutput=1

\documentclass[11pt,twoside,a4paper,cmspaper,final,collab]{cms-tdr}
\def\svnVersion{251224}\def\cmsCernNo{2014-052}\def\cmsCernDate{\today}\def\cmsMessage{Published in Physics Letters B as \href{http://dx.doi.org/10.1016/j.physletb.2014.06.076}{\doi{10.1016/j.physletb.2014.06.076.}}}

\begin{document}\cmsNoteHeader{TOP-12-035}

\hyphenation{had-ron-i-za-tion}
\hyphenation{cal-or-i-me-ter}
\hyphenation{de-vices}

\RCS$Revision: 245072 $
\RCS$HeadURL: svn+ssh://svn.cern.ch/reps/tdr2/papers/TOP-12-035/trunk/TOP-12-035.tex $
\RCS$Id: TOP-12-035.tex 245072 2014-06-05 16:33:45Z alverson $
\newlength\cmsFigWidth
\ifthenelse{\boolean{cms@external}}{\setlength\cmsFigWidth{0.95\columnwidth}}{\setlength\cmsFigWidth{0.6\textwidth}}
\newlength\cmsTabWidth
\ifthenelse{\boolean{cms@external}}{\setlength\cmsTabWidth{\columnwidth}}{\setlength\cmsTabWidth{0.6\textwidth}}
\providecommand{\mt}{\ensuremath{M_\mathrm{T}}\xspace}
\providecommand{\mtop}{\ensuremath{M_\mathrm{t}}\xspace}
\cmsNoteHeader{TOP-12-035} 
\title{Measurement of the ratio $\mathcal{B}(\cPqt\to\PW\cPqb)/\mathcal{B}(\cPqt\to\PW\Pq)$ in pp collisions at $\sqrt{s}=8$\TeV}

\date{\today}

\abstract{
The ratio of the top-quark branching fractions
$\mathcal{R}=\mathcal{B}(\cPqt\to\PW\cPqb)/\mathcal{B}(\cPqt\to\PW\Pq)$,
where the denominator includes the sum over all down-type quarks (q = b, s, d),
is measured in the $\ttbar$ dilepton final state
with proton-proton collision data at $\sqrt{s}=8$\TeV from an
integrated luminosity of 19.7\fbinv,
collected with the CMS detector.
In order to quantify the purity of the signal sample,
the cross section is measured by fitting the observed jet multiplicity,
thereby constraining the signal and background contributions.
By counting the number of b jets per event, an unconstrained value of
$\mathcal{R}=1.014\pm0.003\stat\pm0.032\syst$ is measured,
in a good agreement with
current precision measurements in electroweak and flavour sectors.
A lower limit $\mathcal{R}>0.955$ at the 95\% confidence level is obtained after
requiring $\mathcal{R}\leq 1$,
and a lower limit on the Cabibbo--Kobayashi--Maskawa matrix element
$\abs{V_{\cPqt\cPqb}}>0.975$ is set at 95\% confidence level.
The result is combined with
a previous CMS measurement of the $t$-channel
single-top-quark
cross section
to
determine the top-quark total decay width,
$\Gamma_{\cPqt}=1.36\pm0.02\stat^{+0.14}_{-0.11}\syst$\GeV.
}

\hypersetup{%
pdfauthor={CMS Collaboration},%
pdftitle={Measurement of the ratio B(t to Wb)/B(t to Wq) in pp collisions at sqrt(s) = 8 TeV},%
pdfsubject={CMS},%
pdfkeywords={CMS, top, CKM, width}}

\maketitle 

\section{Introduction}
\label{sec:intro}

Because of its large mass
\cite{ATLAS:2014wva},
the top quark decays
before fragmenting or forming a hadronic bound state~\cite{Bigi:1986sr}.
According to the standard model (SM), the top quark decays through
an electroweak interaction almost exclusively to an on-shell W~boson and a b~quark.
The magnitude of the top-bottom charged current is proportional to $\abs{V_{\cPqt\cPqb}}$,
an element of the Cabibbo--Kobayashi--Maskawa (CKM) matrix.
Under the assumption that the CKM matrix is unitary
and given the measured values for $V_{\cPqu\cPqb}$ and $V_{\cPqc\cPqb}$
(or $V_{\cPqt\cPqs}$ and $V_{\cPqt\cPqd}$), $\abs{V_{\cPqt\cPqb}}$
is expected to be close to unity and dominate over the off-diagonal elements,
 \ie $\abs{V_{\cPqt\cPqb}} \gg \abs{V_{\cPqt\cPqs}},\abs{V_{\cPqt\cPqd}}$.
Thus, the decay modes of the top quark to lighter down-type quarks (d or s) are allowed, but highly suppressed.
The indirect measurement of $\abs{V_{\cPqt\cPqb}}$, from the
unitarity constraint of the CKM matrix, is $\abs{V_{\cPqt\cPqb}}=0.999146^{+0.000021}_{-0.000046}$~\cite{Beringer:1900zz}.
Any deviation from this value or in the partial decay width of the top quark to $\cPqb$ quarks,
would indicate new physics contributions
such as those from a
new heavy up- and/or down-type quarks or a charged Higgs boson, amongst others~\cite{Alwall:2006bx}.
Direct searches at the Large Hadron Collider (LHC) have set lower limits on the mass of these hypothetical new
particles~\cite{Chatrchyan:2012yea,Chatrchyan:2012fp,Aad:2012bt,CMS:2012ab,ATLAS:2012qe,Aad:2012pga,Chatrchyan:2013uxa,Chatrchyan:2013wfa,Chatrchyan:2012vca,Aad:2013hla,Aad:2012rjx},
and the observation of a SM Higgs boson
candidate~\cite{oai:arXiv.org:1207.7214,Chatrchyan:2012plb,Chatrchyan:2013lba}
places stringent constraints on the existence of a fourth
sequential generation of quarks.
These results support the validity of
both the unitarity hypothesis and the $3\times3$ structure of the CKM matrix
for the energy scale probed by the LHC experiments.
However, other new physics contributions, including those described above,
could invalidate the bounds established so far on $\abs{V_{\cPqt\cPqb}}$~\cite{Beringer:1900zz}.

In this Letter, we present a measurement of
$\mathcal{R}=\mathcal{B}(\cPqt\to\PW\cPqb)/\mathcal{B}(\cPqt\to\PW\Pq)$,
where the denominator includes the sum over the
branching fractions of the top quark to a $\PW$ boson and a
down-type quark (q = b, s, d).
Under the assumption of the unitarity of the $3\times3$ CKM matrix,
$\mathcal{R}=\abs{V_{\cPqt\cPqb}}^2$,
and thus to indirectly measure  $\abs{V_{\cPqt\cPqb}}$.
In addition, the combination of a determination of $\mathcal{R}$ and a
measurement
of the $t$-channel single-top cross section can
provide an indirect measurement of the top-quark width ($\Gamma_{\cPqt}$)~\cite{Yuan:1989tc}.
The most recent measurement of $\Gamma_{\cPqt}$ based on this approach~\cite{Abazov:2012vd}
is found to be compatible with the SM predictions
with a relative uncertainty of approximately  22\%.
The value of $\mathcal{R}$ has been measured at the Tevatron, and the
most precise result is obtained by the D0 Collaboration, where
$\mathcal{R} = 0.90 \pm 0.04\,\text{(stat.+syst.)}$~\cite{Abazov:2011zk}
indicates a tension with the SM prediction.
This tension is enhanced for the measurement in the \ttbar dilepton decay channel, where both $\PW$ bosons decay leptonically
and $\mathcal{R} = 0.86^{+0.041}_{-0.042}\stat\pm{0.035}\syst$ is obtained.
The most recent measurements by the CDF Collaboration
are given in~\cite{Aaltonen:2013doa,Aaltonen:2014yua}.

Owing to its purity, the \ttbar dilepton channel is chosen for this measurement.
Events are selected from the data sample acquired in proton-proton collisions at $\sqrt{s}=8$\TeV
by the Compact Muon Solenoid (CMS) experiment at the LHC during 2012.
The integrated luminosity of the analyzed data sample is $19.7\pm0.5$\fbinv~\cite{CMS-PAS-LUM-13-001}.
The selected events are used to measure the \ttbar production cross
section by fitting the observed jet multiplicity distribution,
constraining the signal and background contributions.
The $\cPqb$-quark content of the events is inferred from the distribution
of the number of $\cPqb$-tagged jets per event
as a function of jet multiplicity for each of the dilepton channels.
Data-based strategies are used to constrain
the main backgrounds
and the contributions of extra jets from gluon radiation in \ttbar events.
The $\mathcal{R}$ value is measured
by fitting the observed $\cPqb$-tagged jet distribution
with a parametric model
that depends on the observed cross section, correcting for the fraction of jets
that cannot be matched to a $\cPqt\to \PW\cPq$ decay.
The model also depends
on the efficiency for identifying $\cPqb$ jets
and discriminating them from other jets.
Lastly, the measurement of $\mathcal{R}$ is combined
with a previously published CMS result of the
 $t$-channel
production cross section
of single top quarks in pp collisions
~\cite{Chatrchyan:2012ep}
to yield an indirect determination
of the top-quark total decay width.

\section{The CMS detector}
\label{sec:cmsexperiment}

The central feature of the Compact Muon Solenoid (CMS) apparatus is a
superconducting solenoid of 6\unit{m} internal diameter,
providing a magnetic field of 3.8\unit{T}. Within the superconducting
solenoid volume are a silicon pixel and strip tracker,
a lead tungstate crystal electromagnetic calorimeter (ECAL),
and a brass/scintillator hadron calorimeter (HCAL),
each composed of a barrel and two endcap sections.
Muons are measured in gas-ionization detectors embedded in the steel
flux-return yoke outside the solenoid.
Extensive forward calorimetry complements the coverage provided by the barrel and endcap detectors.

The silicon tracker measures charged particles within the
pseudorapidity range $\abs{\eta}< 2.5$,
where the pseudorapidity $\eta$ is defined as $\eta = -\ln\left[\tan\left(\theta/2\right)\right]$ and $\theta$ is the polar angle
of the trajectory of the particle with respect to the anticlockwise-beam direction.
The tracker consists of 1440 silicon pixel and 15\:148 silicon strip
detector modules and is located in the
field of the superconducting solenoid. It provides an impact parameter
resolution of 
${\sim}15\mum$
and a transverse momentum (\pt) resolution of about 1.5\% for 100\GeV particles.
The electron energy is measured by the ECAL and its direction is measured by the tracker.
The mass resolution for $\cPZ \to \Pe \Pe$ decays is 1.6\% when both electrons are in the
ECAL barrel, and 2.6\% when both electrons are in the ECAL endcap~\cite{Chatrchyan:2013dga}.
Matching muons to tracks measured in the silicon tracker results
in a \pt resolution between 1 and 10\%, for \pt values up to 1\TeV.
The jet energy resolution (JER) amounts typically to 15\% at 10\GeV, 8\% at
100\GeV, and 4\% at 1\TeV~\cite{Chatrchyan:2011ds}.

A more detailed description of the detector can be found in Ref.~\cite{Chatrchyan:2008zzk}.

\section{Simulation of signal and background events}
\label{sec:simulation}

The top-quark pair production cross section has been calculated
at next-to-next-to-leading order (NNLO) and next-to-next-to-leading
logarithmic soft gluon terms (NNLL)
~\cite{Czakon:2013goa}.
In proton-proton collisions at $\sqrt{s}=8$\TeV, and for a top-quark mass of 172.5\GeV,
the expected cross section is $\sigma_{\rm NNLO+NNLL}(\ttbar)=253~^{+6}_{-8}\,\text{(scale)}\,\pm 6\,\text{(PDF)}\unit{pb}$,
where the first uncertainty is from the factorisation and
renormalisation scales, and the second is from the parton
distribution functions (PDFs).
Signal events are simulated for a top-quark mass of 172.5\GeV with the
leading-order (LO) Monte Carlo (MC) generator
\MADGRAPH (v5.1.3.30)~\cite{Alwall:2011uj}
matched to \PYTHIA (v6.426)~\cite{Sjostrand:2006za},
where the $\tau$ lepton decays are simulated with the
\TAUOLA package (v27.121.5)~\cite{Davidson:2010rw}.
The CTEQ6L1 PDF set is used in the event generation~\cite{Pumplin:2002vw}.
Matrix elements describing up to three partons, and including $\cPqb$ quarks, in addition to the
\ttbar pair are included in the generator used to produce the
simulated signal samples.
An alternative simulation at next-to-leading order (NLO)
based in \POWHEG (v1.0,r1380)~\cite{Nason:2004rx,Frixione:2007vw,Alioli:2010xd},
using the CTEQ6M PDF set~\cite{Pumplin:2002vw} and interfaced with \PYTHIA,
is used to evaluate the signal description uncertainty.
A correction to the simulated top-quark \pt
is applied, based on the approximate NNLO computation~\cite{Kidonakis:2012db}:
the events are reweighted at the generator level to match the top-quark \pt
prediction, and the full difference between
the reweighted and unweighted simulations is assigned as a systematic uncertainty.

The most relevant background processes for the dilepton channel
are from the production
of two genuine isolated leptons with large \pt.
This includes
Drell--Yan (DY)  production of charged leptons, \ie from a $\cPZ/\gamma^{*}$ decay, which
is modelled with \MADGRAPH for dilepton invariant masses above 10\GeV,
and it is normalised to a NNLO cross section of 4.393\unit{nb}, computed
using
\textsc{fewz}~\cite{Melnikov:2006kv}.
The $\cPZ+\gamma$ process is also simulated with \MADGRAPH and
normalized to the LO predicted cross section of 123.9\unit{pb}.
Single-top-quark processes are modelled at NLO
with \POWHEG~\cite{Alioli:2009je,Re:2010bp}
and normalised to cross sections of $22\pm2\unit{pb}$, $86\pm3\unit{pb}$, and $5.6\pm0.2\unit{pb}$ for
the tW, $t$-, and $s$- channel production, respectively~\cite{Kidonakis:2012db}.
The theory uncertainties are due to the variation of the PDFs and factorisation and
renormalisation scales.
Diboson processes are modelled with \MADGRAPH, and normalised to the
NLO cross section computed with \MCFM~\cite{Campbell:2010ff}.
The generation of $\PW\PW$, $\PW\cPZ$, and $\cPZ\cPZ$ pairs is
normalised to inclusive cross sections of 54.8\unit{pb}, 33.2\unit{pb}, and 17.7\unit{pb}, respectively.
For $\PW\cPZ$ and $\cPZ\cPZ$ pairs a minimum dilepton invariant mass of 12\GeV is required.
Associated production of $\PW$ or $\cPZ$ bosons with \ttbar pairs is
modelled with \MADGRAPH, and normalized to the LO cross sections of
232\unit{fb} and 208\unit{fb}, respectively.
The production of a $\PW$ boson in association with jets, which includes misreconstructed and non-prompt leptons, is
modelled with \MADGRAPH and normalised to a total cross section of 36.3\unit{nb} computed with \textsc{fewz}.
Multijet processes are also studied in simulation but are found to yield negligible contributions to the selected sample.

A detector simulation based on \GEANTfour (v.9.4p03)~\cite{Allison:2006ve,Agostinelli:2002hh}
is applied after the generator step for both signal and background samples.
The presence of multiple interactions (pileup) per bunch crossing is incorporated by simulating additional interactions
(both in-time and out-of-time with the collision) with a multiplicity
matching that observed in the data.
The average number of pileup events in the data is 21 interactions per bunch crossing.

\section{Event selection and background determination}
\label{sec:evsel}

The event selection is optimised for  \ttbar dilepton final states
that contain two isolated oppositely charged leptons $\ell$ (electrons or
muons), missing transverse energy (\MET) defined below, and at least two jets.
Events in which the electrons or muons are from
intermediate $\tau$ lepton decays are considered as
signal events.
Dilepton triggers are used to acquire the data samples,
where a minimum transverse momentum of 8\GeV is required for each of the leptons,
and 17\GeV is required for at least one of the leptons.
Electron-based triggers include additional isolation requirements, both in the tracker and calorimeter detectors.

All objects in the events are reconstructed with a particle-flow (PF) algorithm~\cite{CMS-PAS-PFT-09-001,CMS-PAS-PFT-10-001}.
Reconstructed electron and muon candidates
are required to have $\pt>20\GeV$ and to be in the fiducial region
$\abs{\eta}\leq2.4$ of the detector.
A particle-based relative isolation
parameter is computed for each lepton and
corrected on an event-by-event basis for the contribution from pileup events.
We require that the scalar sum of the \pt of all particle candidates
reconstructed in an isolation cone built around the lepton's momentum vector is
less than 15\% (12\%) of the electron (muon) transverse momentum.
The isolation cone is defined using the radius $R=\sqrt{\smash[b]{(\Delta \eta)^{2}+(\Delta \phi)^{2}}}=0.4$,
where $\Delta\eta$ and $\Delta\phi$ are the differences in
pseudorapidity and azimuthal angle between the particle candidate and the lepton.
For each event we require at least two lepton candidates originating
from a single primary vertex.
Among the vertices identified in the event, the vertex
with the largest $\sum \pt^2$, where the sum runs over all
tracks associated with the vertex, is chosen as the primary vertex.
The two leptons with highest $\pt$ are chosen to form the dilepton pair.
Same-flavour dilepton pairs (ee or $\mu\mu$) compatible with $\cPZ\to\ell\ell$ decays are removed
by requiring $\abs{M_\cPZ-M_{\ell\ell}}>15\GeV$, where $M_\cPZ$
is the $\cPZ$ boson mass~\cite{Beringer:1900zz} and
$M_{\ell\ell}$ is the invariant mass of the dilepton system.
For all dilepton channels it is further required that
$M_{\ell\ell}>12\GeV$ in order to veto low-mass dilepton resonances,
and that the leptons have opposite electric charge.

Jets are reconstructed by clustering all the PF candidates
using the anti-\kt algorithm~\cite{Cacciari:2008gp} with a distance parameter of 0.5.
Jet momentum is defined as the vector sum of all particle momenta in the jet,  and in the simulation it is found
 to be within 5 to 10\% of the hadron-level momentum over the entire \pt\ spectrum and detector
acceptance. A correction is applied by subtracting the extra energy clustered in jets due to pileup,
following the procedure  described in Refs.~\cite{Cacciari:2008gn,Cacciari:2007fd}.
The energies of charged-particle candidates associated with other reconstructed primary vertices in the event are also subtracted.
Jet energy scale (JES) corrections are derived from simulation, and are validated with \textit{in-situ}
measurements of the energy balance of dijet and photon+jet events~\cite{Chatrchyan:2011ds}.
Additional selection criteria are applied to events to remove spurious jet-like features originating from
isolated noise patterns in certain HCAL regions.
In the selection of \ttbar events,
at least two jets, each with a corrected transverse momentum $\pt >30\GeV$ and $\abs{\eta}\leq 2.4$, are required.
The jets must be separated from the selected leptons by
$\Delta R(\ell, \text{jet})=\sqrt{\smash[b]{(\Delta\eta)^2+(\Delta\phi)^2}}\geq0.3$.
Events with up to four jets, selected under these criteria, are used.

The magnitude of the vector sum of the transverse momenta of all particles reconstructed in the event
is used as the estimator for the momentum imbalance in the transverse plane, \MET.
All JES corrections applied to the event are also propagated into the \MET estimate.
For the ee and $\mu\mu$ channels, $\MET>40\GeV$ is required in order to reduce the contamination
from lepton pairs produced through the DY mechanism in association with at least two jets.

The DY contribution to the same-flavour dilepton channels is estimated from the data after the full event selection
through the modelling of the angle $\Theta_{\ell\ell}$ between the two leptons.
The $\Theta_{\ell\ell}$ distribution discriminates between leptons produced in DY processes
and leptons from the top-quark pair decay cascade.
In the first case an angular correlation is expected, while in the
second case the leptons are nearly uncorrelated.
The probability distribution function for $\Theta_{\ell\ell}$
is derived from data using a DY-enriched control region selected after inverting the \MET requirement
of the standard selection. Studies of simulated events indicate that
the shape of the
$\Theta_{\ell\ell}$ distribution is well described with this method, and that the contamination from other processes
in the control region can be neglected. Compatibility tests performed in simulations using different channels and
jet multiplicities are used to estimate an intrinsic 10\% uncertainty in the final DY background.
The other sources of uncertainty in the method are related to the simulation-based
description of the probability distribution function for the
$\Theta_{\ell\ell}$ distribution from other processes.
Uncertainties are estimated either by propagating the uncertainties in pileup or JES and JER,
or by trying alternative functions for the \ttbar contribution with varied
factorisation/renormalisation scales ($\mu_\mathrm{R}/\mu_\mathrm{F}$)
with respect to their nominal values given by the momentum transfer in
the event, matrix element/parton shower (ME-PS) matching threshold, or generator choice (\POWHEG vs. \MADGRAPH).
The shapes of kinematic distributions for DY and other processes are used in a maximum-likelihood fit to estimate the amount of DY
background in the selected sample.
A total uncertainty of 21\% is estimated from the data in the rate of DY events for the same-flavour channels.

For the $\Pe\Pgm$ channel, a similar fit procedure is adopted using a different variable: the transverse mass $\mt=\sqrt{\smash[b]{2\MET\pt(1-\cos\Delta\phi)}}$
of each lepton, where $\Delta\phi$ is the difference in azimuthal
angle between the lepton and the missing transverse momentum.
The distribution of the sum $\sum \mt$ is used as the
distribution in the fit.
In this case the probability distribution function for
$\cPZ/\gamma^{*}\to \tau\tau\to \Pe\mu$ is derived from simulation.
The determination of the uncertainty associated with this method follows
a similar prescription to that described above for the same-flavour channels.
A total uncertainty of 21\% is assigned to the amount of
DY contamination in the $\Pe\Pgm$ channel.

The second-largest background contribution is from single-top-quark processes
(in particular the tW channel) that is relevant for this measurement since the decay products
of a single top quark (instead of a pair) are selected.
The contribution of this process is estimated from simulation.
Other background processes are also estimated from simulation.
Uncertainties in the normalisation stemming from instrumental uncertainties
in the integrated luminosity, trigger and selection efficiencies, and energy scales,
as well as generator-specific uncertainties, are taken into account.

Table~\ref{tab:summaryyields} shows the yields in the data and those predicted for signal and background events
after the full event selection.
The systematic uncertainties assigned to the predictions of signal and background events include the uncertainties
in the JES and JER, pileup modelling,
cross section calculations, integrated luminosity, and trigger and selection
efficiencies.
A conservative uncertainty is assigned to the predicted yields of multijet and
$\PW\to\ell\nu$ background events since these contributions are from
misidentified leptons and have been estimated solely from simulation.
Good overall agreement is observed
for all three dilepton categories between the yields in data and the
sum of expected yields.

\begin{table*}
\topcaption{Predicted and observed event yields after the full event selection.
The combination of statistical uncertainties with experimental and theoretical systematic uncertainties is reported.
Non dileptonic \ttbar channels, identified using a generator-level matching,
as well as associated production with vector bosons
($\PW$ or $\cPZ$), is designated as ``Other \ttbar'' and grouped with
the expected contribution from single $\PW$ boson and multijets productions.
The expected contribution from vector boson pair processes is designated as ``VV''.
}
\label{tab:summaryyields}
\centering
\begin{tabular}{lccc} \hline
Source                              & $\Pe\Pe$                  & $\mu\mu$ & $\Pe\Pgm$ \\\hline
$\PW\to \ell\nu$ , multijets, other \ttbar   & $134\pm91$ & $43\pm10$& $(38\pm20)\times 10$  \\
VV                                     & $292\pm15$ & $333\pm16$ & $995\pm39$ \\
$\cPZ/\gamma^{*}\to \ell\ell$ & $(297\pm63)\times 10$ & $(374\pm79)\times 10$ & $(184\pm39)\times 10$  \\
Single top quark              & $526\pm26$ &$583\pm26$ & $1834\pm64$ \\
\ttbar dileptons (signal) & $(1003\pm50)\times 10$ & $(1104\pm54)\times 10$ & $(349\pm17)\times 10^2$ \\\hline
Total                                 & $(1395\pm81)\times 10$ & $(1574\pm96)\times 10$ & $(400\pm17)\times 10^2$ \\\hline
Data                                 & $13723$            & $15596$ & $38892$ \\
\hline
\end{tabular}
\end{table*}

\section{Cross section measurement}
\label{sec:samplecomposition}

\begin{figure}[htbp]
\centering
\includegraphics[width=0.48\textwidth]{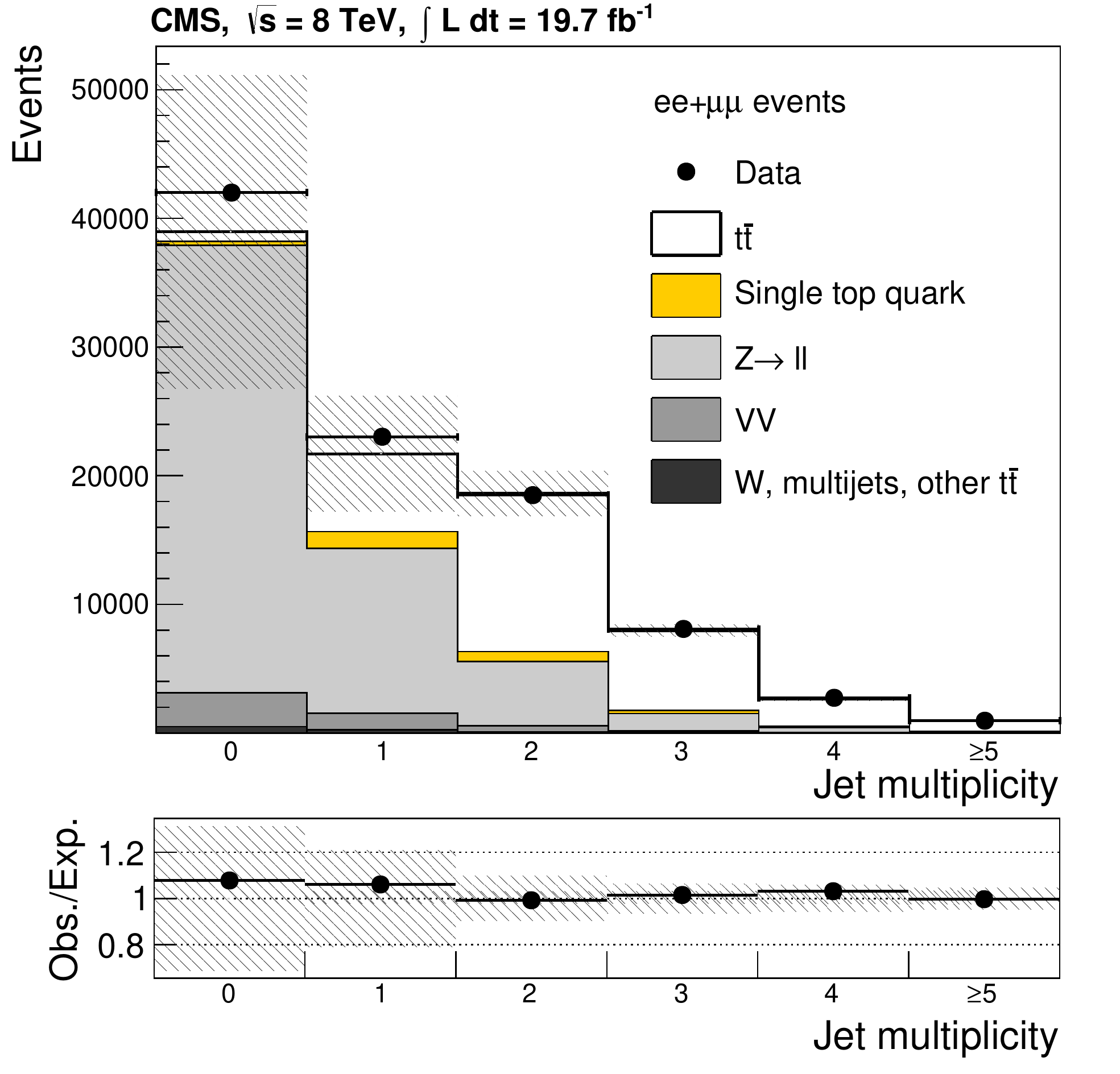}
\includegraphics[width=0.48\textwidth]{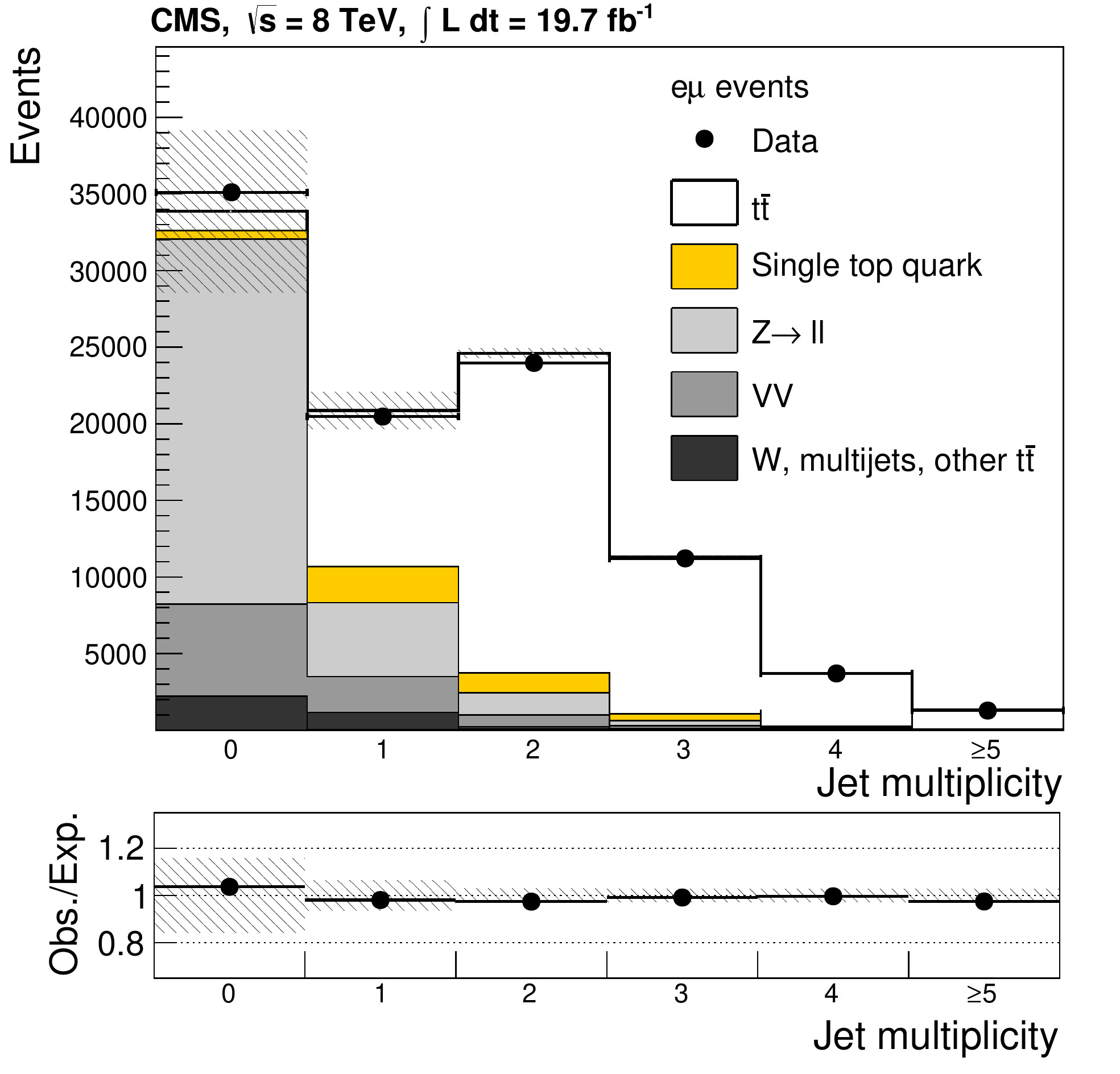}
\caption{
The upper plots show the
observed jet multiplicity after the full event selection,
  except for the requirement on the
number of jets, in the same-flavour (\textit{top}) and different-flavour (\textit{bottom}) channels.
The expectations are shown as stacked histograms, while the
observed data distributions are represented as closed circles.
The predicted distributions for the simulated \ttbar and single-top-quark events correspond to a scenario with $\mathcal{R}=1$.
The lower panels show the ratio of the data to the expectations.
The shaded bands represent the systematic uncertainty in the determination of the main background (DY)
and the integrated luminosity, and vary from 31\% (16\%) to 5\%
(3\%) in the  same- (different-) flavour channels when going from the
$0$ jets to $\geq 5$ jets bin.}
\label{fig:jetmult}
\end{figure}

The selected events are categorized by the dilepton channel and the
number of observed jets.
Figure~\ref{fig:jetmult} shows the expected composition for each event category.
Good agreement is observed between the distributions from the data and
the expectations, including the control regions, defined as events
with fewer than two or more than four jets. The chosen categorization
not only allows one to study
the contamination from initial- and final-state gluon radiation (ISR/FSR) in the sample, but also to
constrain some of the uncertainties from the data.

The \ttbar dilepton signal strength, $\mu$, defined as the ratio of the observed to
the expected signal rate, is measured from the jet multiplicity distribution by using a  profile likelihood method~\cite{Cowan:2010js}.
A likelihood is calculated from the observed number of events in the
$k$ 
dilepton channels and jet multiplicity
categories as
\begin{equation}
\mathcal{L}(\mu,\theta) =
\prod_k \mathcal{P} \left[ N_k, \hat{N}_k(\mu,\theta_i) \right]\cdot \prod_i \rho(\theta_i),
\label{eq:xsecll}
\end{equation}
where $\mathcal{P}$ is the Poisson probability density function,
$N_k$ is the number of events observed in the $k$-th category,
$\hat{N}_k$ is the total number of expected events from signal and background,
and $\theta_i$ are the nuisance parameters, distributed according to a probability density function $\rho$.
The nuisance parameters are used to modify the expected
number of events according to the different systematic uncertainty sources, which include instrumental effects
(such as integrated luminosity, pileup, energy scale and resolution, lepton trigger and selection efficiencies)
and signal modelling ($\mu_\mathrm{R}/\mu_\mathrm{F}$, ME-PS scale, top-quark mass, leptonic branching fractions
of the $\PW$ boson) amongst others.
The PDF uncertainty is estimated using the PDF4LHC prescription~\cite{Alekhin:2011sk,Botje:2011sn}.
The uncertainty from the choice of the \ttbar signal generator is estimated by
assigning the difference between the \MADGRAPH-based and the
\POWHEG-based predictions as an extra uncertainty in the fit.
The nuisance parameters are assumed to be unbiased
and distributed according to a log-normal function.
Based on the likelihood expressed in Eq.~(\ref{eq:xsecll}), the profile
likelihood ratio (PLR) $\lambda$ is defined as
\begin{equation}
\lambda(\mu) = \frac{\mathcal{L}(\mu,\hat{\hat\theta})}{\mathcal{L}(\hat{\mu},\hat\theta)},
\label{eq:profll}
\end{equation}
where the denominator has estimators $\hat{\mu}$ and
$\hat\theta$ that maximise the likelihood, and
the numerator has estimators $\hat{\hat\theta}$ that maximise
the likelihood for the specified signal strength $\mu$.
The signal strength is obtained after maximising $\lambda(\mu)$ in Eq.~(\ref{eq:profll}).
This approach allows us to parameterise the effect of the systematic
uncertainties in the fit.

The signal strength $\mu$ is determined independently in each category,
\ie for each dilepton channel and jet multiplicity. For each category,
the purity of the selected sample ($f_{\ttbar}$) is defined as
the fraction of  ``true" \ttbar signal events in the selected sample,
$f_{\ttbar}=\mu\cdot N_{\text{\ttbar exp}}/N_\text{obs}$,
where $N_{\text{\ttbar exp}}$ is the number of expected \ttbar events,
and $N_\text{obs}$ is the total number of observed events.
By performing the fit for each category, the purity of
the sample is obtained.
The results are summarized in Table~\ref{tab:samplepurity}.
As expected, the $\Pe\mu$ category has the highest purity
($\approx$90\%). Because of the contamination from DY events,
the same-flavour channels have lower purity ($\approx$70\%).
Overall, the signal purity increases with higher jet multiplicity.

As a cross-check, a fit including all
categories, gives the range $0.909<\mu<1.043$ at the 68\% confidence level (CL).
This leads to a \ttbar production cross section of
\begin{equation*}
\sigma(\ttbar)=238\pm1\stat\pm15\syst\unit{pb},
\end{equation*}
in good agreement with NNLO+NNLL expectation~\cite{Czakon:2013goa}
and the latest CMS measurement~\cite{Chatrchyan:2013faa}.
The result is also found to be consistent with the individual results
obtained in each event category.
An extra uncertainty is assigned in the extrapolation of the cross section to the full
phase space because of the dependence of the acceptance
on $\mu_\mathrm{R}/\mu_\mathrm{F}$, ME-PS threshold choices, and the top-quark mass.

The relative single-top-quark contribution ($k_\mathrm{st}$), defined as the ratio of the expected
number of single-top-quark events to the estimated number of inclusive
\ttbar events, is also shown
in Table~\ref{tab:samplepurity} for each category.
For this determination we use the expected number of single-top-quark events
obtained after maximising the PLR in Eq.~(\ref{eq:profll}).
The contribution due to single-top-quark events tends to be most significant in the two-jet category ($<7\%$
relative to inclusive \ttbar events).
Since the estimate is obtained for a specific scenario in which $\mathcal{R}=1$,
an extra linear dependence of $k_\mathrm{st}$ on $\mathcal{R}$ is introduced
in order to account for
the increase in the t\PW\ cross section
as $\abs{V_{\cPqt\cPqb}}$ becomes smaller while
$\abs{V_{\cPqt\cPqd}}$ and $\abs{V_{\cPqt\cPqs}}$ become larger~\cite{Alwall:2006bx}.
In this parameterisation, the measured ratio $\abs{V_{\cPqt\cPqd}}/\abs{V_{\cPqt\cPqs}}=0.211\pm0.006$ is used~\cite{Beringer:1900zz},
and the uncertainty
is considered as an intrinsic systematic uncertainty in the measurement of $\mathcal{R}$.

\begin{table*}
\topcaption{
Fraction of \ttbar events ($f_{\ttbar}$) and relative
contribution from single-top-quark processes
($k_\mathrm{st}$) for various jet multiplicities and dilepton channels, as determined from the
profile likelihood fit. The total uncertainty is shown.
}
\label{tab:samplepurity}
\centering
\begin{tabular}{cccccc}\hline
\multirow{2}{*}{Parameter} & \multirow{2}{*}{Jet multiplicity}  & \multicolumn{3}{c}{Dilepton channel} \\
& & $\Pe\Pe$ & $\mu\mu$ & $\Pe\mu$\\\hline
\multirow{3}{*}{$f_{\ttbar}$}
& 2 & $0.67\pm0.07$  & $0.65\pm0.08$ & $0.85\pm0.06$ \\
& 3 & $0.79\pm0.06$  & $0.78\pm0.07$  & $0.90\pm0.07$\\
& 4 & $0.81\pm0.11$  & $0.82\pm0.11$ & $0.94\pm0.10$\\\hline
\multirow{3}{*}{$k_\mathrm{st}$}
& 2 & $0.062\pm0.004$ & $0.063\pm0.004$ & $0.062\pm0.003$\\
& 3 & $0.040\pm0.003$ & $0.040\pm0.003$ & $0.041\pm0.002$ \\
& 4 & $0.036\pm0.004$ & $0.036\pm0.006$ & $0.029\pm0.003$ \\
\hline
\end{tabular}
\end{table*}

\section{Probing the \texorpdfstring{$\cPqb$}{b}-flavour content}
\label{sec:probinghfc}

In this section
the $\cPqb$-flavour content of the selected events (both signal and
background) is determined from the $\cPqb$-tagged jet multiplicity distribution.
The probability of incorrectly assigning a jet
must be evaluated (Section~\ref{subsec:jetassignmentestimate})
in order to correctly estimate the heavy-flavour content of top-quark decays (Section~\ref{subsec:heavyflavourcontentfit}).

The $\cPqb$-tagging algorithm that is used (the combined secondary vertex, CSV method
described in Ref.~\cite{Chatrchyan:2012jua})
is a multivariate procedure in which both information on the
transverse impact parameter with respect to the primary vertex of the associated tracks,
and the reconstructed secondary vertices is used
to discriminate $\cPqb$ jets from c, light-flavour (u, d, s) and gluon jets.
The $\cPqb$-tagging efficiency ($\varepsilon_\cPqb$) is measured~\cite{CMS-PAS-BTV-13-001} using multijet events
where a muon is reconstructed inside a jet;
a data-to-simulation scale factor is derived
and is used to correct the predicted $\varepsilon_\cPqb$ value
in the \ttbar dilepton sample from simulation.
After correction, the expected efficiency in the selected \ttbar sample is $\approx84\%$,
and the uncertainty in the scale factor
from the data is 1--3\%,~depending on the kinematics of the jets~\cite{CMS-PAS-BTV-13-001}.
The same scale factor is applied to the expected $\cPqc$-tagging
efficiency but with a doubled uncertainty with respect to the one
assigned to $\cPqb$ jets
owing to the fact that no direct measurement of the $\cPqc$-tagging
efficiency is performed.
For jets originating from the hadronisation of light-flavour jets, the
misidentification efficiency ($\varepsilon_\Pq$)
is evaluated~\cite{Chatrchyan:2012jua} from so-called negative tags in jet samples,
which are selected using tracks that have a negative impact parameter or
secondary vertices with a negative decay length.
The scalar product of the jet direction with the vector pointing from the primary vertex to
the point of closest approach of a track with negative impact
parameter has the opposite sign of the scalar product  taken with
respect to the point of closest approach.
The data-to-simulation correction factor for the misidentification efficiency
is known with an uncertainty of
about 11\%,
and the expected misidentification efficiency in the selected sample is approximately 12\%~\cite{CMS-PAS-BTV-13-001}.

\begin{figure*}[htb]
\centering
\includegraphics[width=0.99\textwidth]{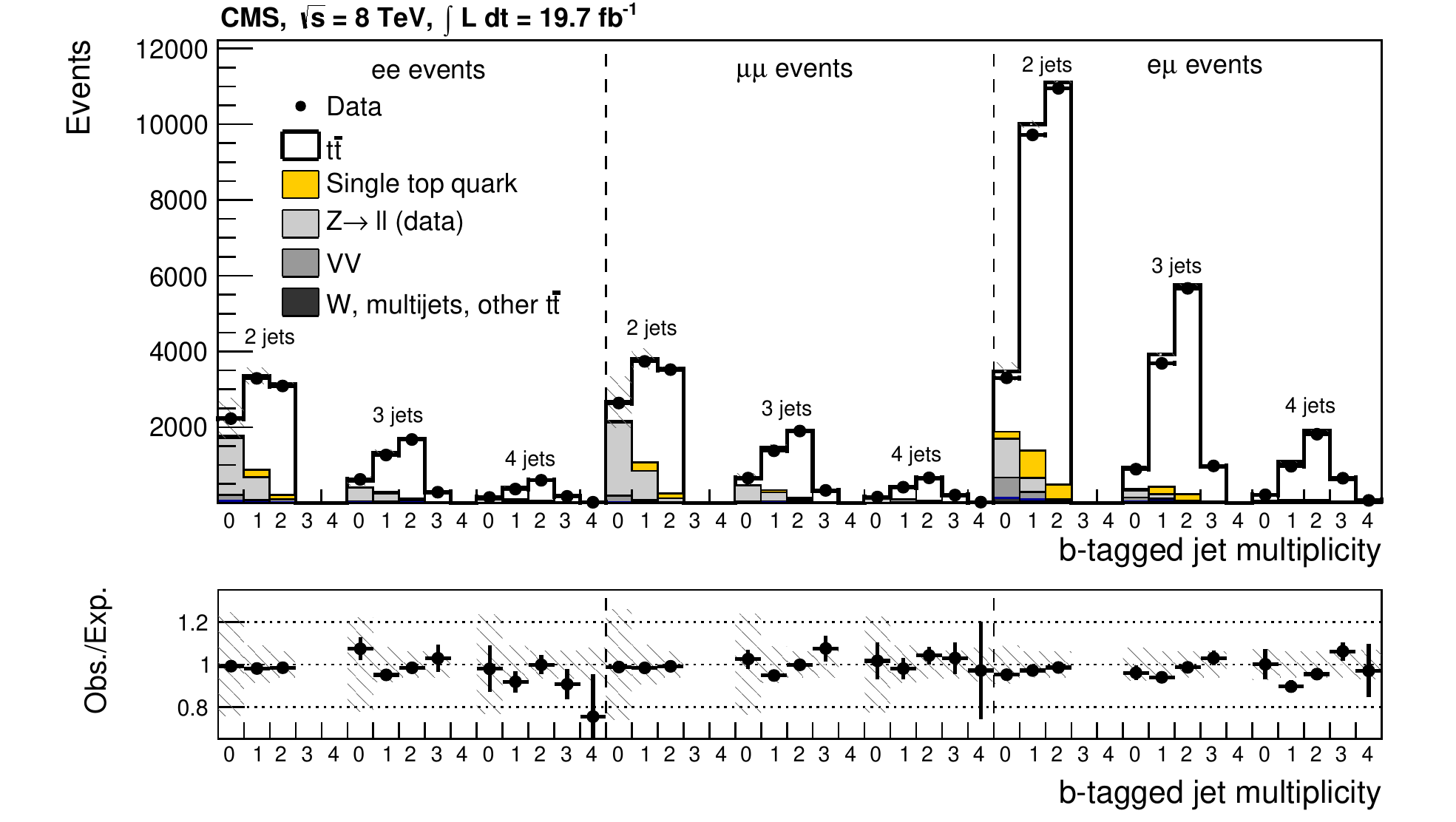}
\caption{The upper plot shows the number of $\cPqb$-tagged jets per event for the different \ttbar dilepton channels.
For each final state, separate subsets are shown corresponding to events with two, three, or four jets.
The simulated \ttbar and single-top-quark events correspond to a scenario with $\mathcal{R}=1$.
The lower panel shows the ratio of the data to the expectations.
The shaded bands represent the uncertainty owing to the finite size of
the simulation samples,
the main background contribution (DY), and the integrated luminosity.}
\label{fig:btagmult}
\end{figure*}

Figure~\ref{fig:btagmult} shows the number of $\cPqb$-tagged jets
in the selected dilepton data sample,
compared to the expectations from simulation.
The multiplicity is shown separately for each dilepton channel and jet multiplicity.
The expected event yields are corrected after the PLR fit for the signal strength
(described in the previous section) and also incorporate the data-to-simulation
scale factors for $\varepsilon_\cPqb$ and $\varepsilon_\Pq$.
Data and simulation agree within 5\%.
The residual differences can be related to the
different number of jets selected from top-quark decays in data and
simulation, the modelling of gluon radiation (ISR/FSR)
and if $\mathcal{R}$ is different from unity, (which is an assumption
used in the simulation).

\subsection{Jet misassignment}
\label{subsec:jetassignmentestimate}

There is a non-negligible probability that at least one jet from a
\ttbar decay is missed, either because
it falls outside of the detector acceptance or is not reconstructed, and another
jet from a radiative process is chosen instead.
In the following discussion, this is referred to as a ``misassigned jet''.
Conversely, jets that come from a top-quark decay will be referred to as ``correctly assigned''.
The rate of correct jet assignments is estimated from the data
using a combination of three different categories:

\begin{itemize}
\item events with no jets selected from top-quark decays,
which also includes background events with no top quarks;
\item events with only one jet from a top-quark decay, which includes
  some \ttbar events and single-top-quark events (mainly produced through the t\PW\ channel);
\item events with two jets produced from the two top-quark decays.
\end{itemize}

\begin{figure}[htp]
\centering
\includegraphics[width=0.48\textwidth]{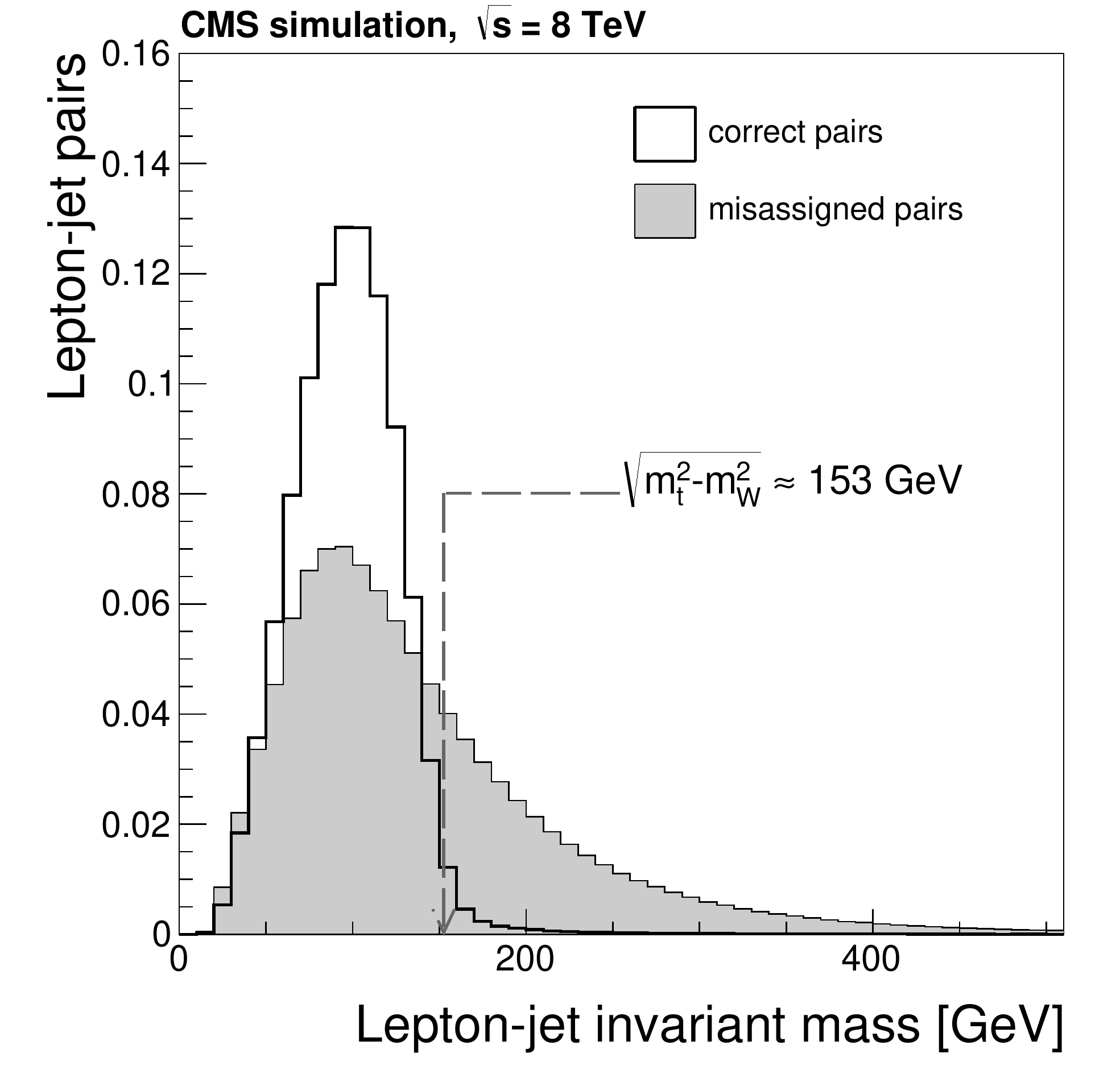}
\includegraphics[width=0.48\textwidth]{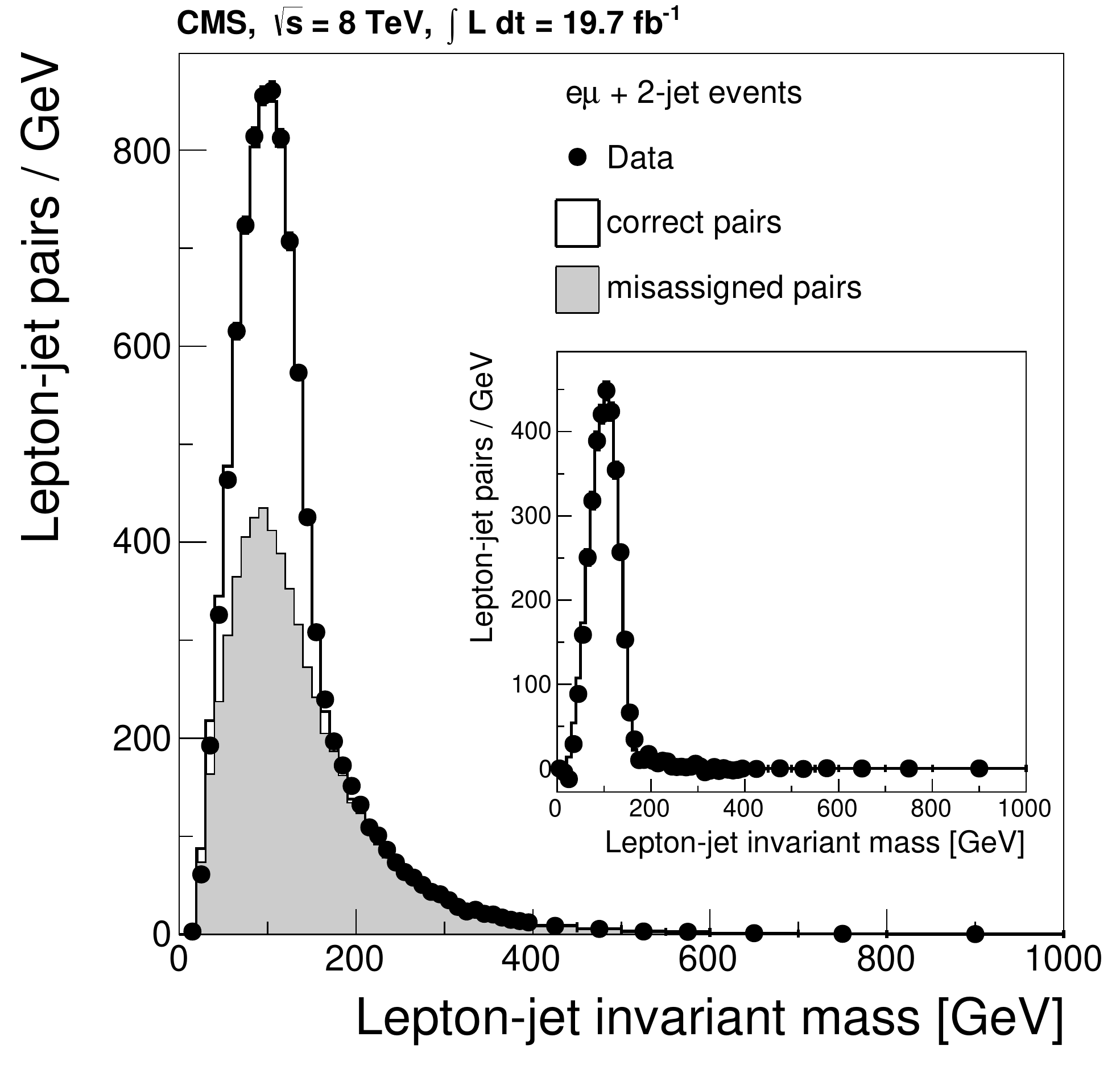}
\caption{
The top plot shows the correct and misassigned lepton-jet
invariant-mass spectra
in simulated \ttbar dilepton events. Both distributions are normalised
to unity.
The endpoint of the spectrum for correctly assigned pairs is shown by the dashed line.
In the bottom plot the observed data is compared with the correct (from simulation) and misassigned (from the data)
components for the lepton-jet invariant-mass spectra in $\Pe\Pgm$ events with exactly two jets.
The lepton-jet mass distribution is shown in the inset,
after the misassigned pairs are subtracted.
}
\label{fig:mlj}
\end{figure}

In order to avoid model uncertainties,
the number of selected jets from top-quark decays is derived from
the lepton-jet invariant-mass ($M_{\ell\mathrm{j}}$) distribution, reconstructed by pairing each lepton with all selected jets.
For lepton-jet pairs originating from the same top-quark decay,  the endpoint of the spectrum occurs at
$M_{\ell\mathrm{j}}\approx\sqrt{\smash[b]{\mtop^2-M_\PW^2}}\approx153\GeV$~\cite{Ellis:1991qj},
where $\mtop$ ($M_{\PW}$) is the top-quark (W boson) mass
(Fig.~\ref{fig:mlj}, \textit{top}, open histogram).
The predicted distribution for correct pairings is obtained after matching the simulated
reconstructed jets to the $\cPqb$ quarks from $\cPqt\to\PW\cPqb$ at the generator
level using a cone of radius $R=0.3$.
The same quantity calculated for a lepton from a top-quark decay paired with
a jet from the top antiquark decay and vice versa (``wrong" pairing)
shows a distribution with a long tail
(Fig.~\ref{fig:mlj}, \textit{top}, filled histogram),
which can be used as a discriminating feature.
A similar tail is observed for ``unmatched'' pairings:
either background processes without top quarks, or leptons matched to other jets.
The combinations with $M_{\ell\mathrm{j}}>180\GeV$ are dominated by
incorrectly paired jets, and this control region is used to normalise the contribution from background.

In order to model the lepton-jet invariant-mass distribution of the misassigned jets,
an empirical method is used based
on the assumption of uncorrelated kinematics.
The validity of the method has been tested using simulation.
For each event in data, the momentum vector of the selected lepton is ``randomly rotated''
with uniform probability in the $\left(\cos\left(\theta\right),\phi\right)$ phase space,
and the $M_{\ell\mathrm{j}}$ is recomputed. This generates a combinatorial distribution that is
used to describe the true distribution of $M_{\ell\mathrm{j}}$ for
misassigned jets.
Figure~\ref{fig:mlj} (\textit{bottom}) compares the data distribution with
the two components of the $M_{\ell\mathrm{j}}$
spectrum, \ie ``correct assignments'' from simulation and ``wrong
assignments'' modelled from the data.
The background model provides a good estimate of the shape of the spectrum of the misassigned
lepton-jet pairs.
After fitting the fractions of the two components to the data,
the  ``misassigned'' contribution is subtracted from the inclusive spectrum,
and the result is compared to the expected
contribution from the correctly assigned lepton-jet pairs.
The result of this procedure is shown in the inset of Fig.~\ref{fig:mlj} (\textit{bottom}).
This method is used to determine the fraction  ($f_\text{correct}$) of selected jets from top-quark decays
in the $M_{\ell\mathrm{j}}$ spectrum.
Consequently, by measuring $f_\text{correct}$, we estimate directly
from the data the number of top-quark decays reconstructed and selected.
Notice that $f_\text{correct}$ cannot be larger than $1/n$ for events
with $n$ jets, as it includes the combinatorial contribution by definition.

In Table~\ref{tab:correctcontributions} the values of $f_\text{correct}$ found in the data are compared to those predicted from simulation.
These include both the contamination from background events
as well as the effect of missing one or two jets from top-quark decays after selection.
The systematic uncertainties affecting the estimate of $f_\text{correct}$ can be split into two sources:

\begin{itemize}
\item distortion of the $M_{\ell\mathrm{j}}$ shape due to the JES and JER of the reconstructed objects~\cite{Chatrchyan:2011ds};
\item calibration uncertainties (derived in the previous section)
  owing to the uncertainty in the $\mu_\mathrm{R}/\mu_\mathrm{F}$ scale,
  the simulation of gluon radiation and the underlying event,
  the top-quark mass value used in simulation,
  and the contributions from background processes.
\end{itemize}

For each case the fit is repeated with different signal probability distribution functions.
The systematic uncertainty is estimated to be 3--10\%, depending on the
jet multiplicity in the event, and is dominated by the ME-PS matching threshold and the
$\mu_\mathrm{R}/\mu_\mathrm{F}$ scale uncertainties.

\begin{table*}
\centering
\topcaption{Fraction of lepton-jet pairs correctly assigned
    in the selected events estimated from the data and predicted from
    simulation. The last column shows the ratio of the fraction
    measured in data to the
    prediction from simulation. The total uncertainty is shown.}
\label{tab:correctcontributions}
\begin{tabular}{ccccc} \hline
Dilepton channel & \# jets & $f_\text{correct}^\text{data}$ & $f_\text{correct}^\mathrm{MC}$ & data/MC \\\hline
\multirow{3}{*}{$\Pe\Pe$}
& 2 & $0.28\pm0.05$ & $0.277\pm0.001$ & $1.03\pm0.19$ \\
& 3 & $0.22\pm0.07$ & $0.223\pm0.001$ & $0.99\pm0.29$ \\
& 4 & $0.19\pm0.07$ & $0.175\pm0.001$ & $1.09\pm0.43$ \\\hline
\multirow{3}{*}{$\mu\mu$}
& 2 & $0.28\pm0.06$ & $0.276\pm0.001$  & $1.00\pm0.21$ \\
& 3 & $0.24\pm0.06$ & $0.227\pm0.001$ & $1.05\pm0.25$ \\
& 4 & $0.20\pm0.07$ & $0.181\pm0.001$ & $1.08\pm0.37$ \\\hline
\multirow{3}{*}{$\Pe\Pgm$}
& 2 & $0.36\pm0.06$ & $0.3577\pm0.0007$ & $1.01\pm0.16$ \\
& 3 & $0.26\pm0.05$ & $0.2625\pm0.0007$ & $1.00\pm0.18$ \\
& 4 & $0.21\pm0.06$ & $0.2047\pm0.0008$ & $1.00\pm0.27$ \\\hline
\end{tabular}
\end{table*}

\begin{figure}[htp]
\centering
\includegraphics[width=\cmsFigWidth]{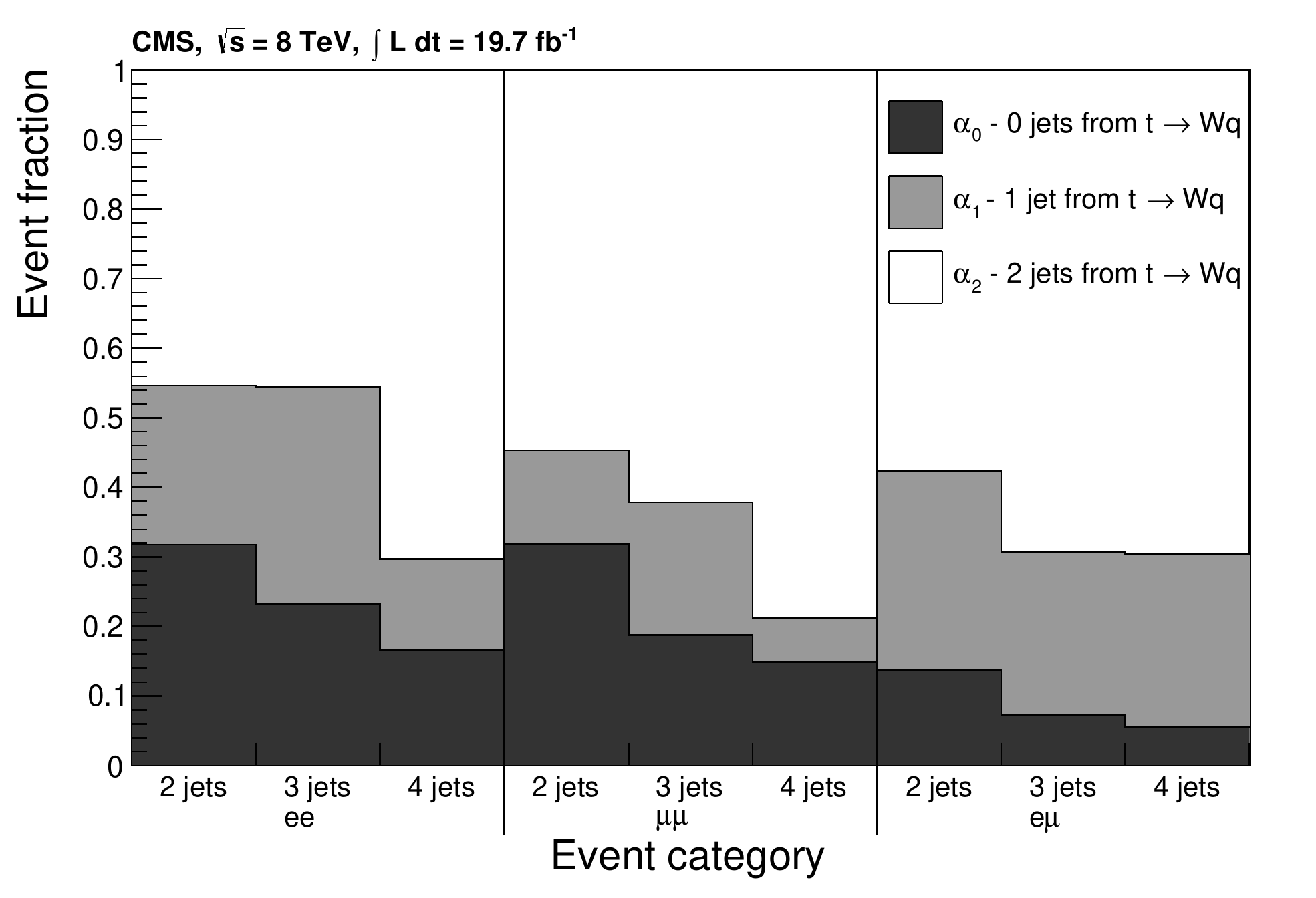}
\caption{
Fraction of events with 0, 1, or 2 top-quark decays selected,
as determined from the data: these fractions, shown for different event categories,
are labeled $\alpha_0$, $\alpha_1$, and $\alpha_2$, respectively.
}
\label{fig:alphaobs}
\end{figure}

By combining the measured $f_\text{correct}$  from the data with the fraction of \ttbar and single-top-quark events,
a parameterisation of the three classes of events is obtained:
\ie the number of events with 0, 1, or 2 selected top-quark decays.
The relative amounts of the three event classes are parameterised by
the probabilities $\alpha_i$, where $i$ corresponds to the number of
jets from top-quark decays selected in an event.
The probabilities $\alpha_i$ are constrained to $\sum_i\alpha_i=1$.
Figure~\ref{fig:alphaobs} summarizes the values of $\alpha_i$ obtained
for the individual event categories,
where the differences are dominated
by the event selection efficiencies and the background
contribution in each category.

\subsection{Heavy-flavour content}
\label{subsec:heavyflavourcontentfit}

For a given number of correctly reconstructed and selected jets, the expected $\cPqb$-tagged jet multiplicity
can be modelled as a function of $\mathcal{R}$ and the $\cPqb$-tagging and misidentification efficiencies.
In the parameterisation, we distinguish events containing jets from 0,
1, or 2 top-quark decays.
The model is an extension of the one proposed in Ref.~\cite{Silva:2010qt}.
For illustration, the most significant case is considered, \ie modelling the
observation of two $\cPqb$-tagged jets in an event with two reconstructed jets.
For the case where two jets from top-quark decays are selected in the event,
the probability to observe two $\cPqb$-tagged jets can be written as
\begin{equation}
P_\mathrm{ 2j,2t,2d}=\mathcal{R}^2\varepsilon_\cPqb^2+2\mathcal{R}(1-\mathcal{R})\varepsilon_\cPqb\varepsilon_\Pq+(1-\mathcal{R})^2\varepsilon_\Pq^2,
\label{eq:prob22}
\end{equation}
where the subscripts (2j, 2t, 2d) indicate a two-jet event, with two
$\cPqb$-tagged jets, and two top-quark decays.
If instead, only one jet from a top-quark decay is present in the event,
the probability is modified to take
the second jet into account in the measurement of $\mathcal{R}$.
In this case, the probability of observing two $\cPqb$-tagged jets is
\begin{equation}
 P_\text{2j,2t,1d}=\mathcal{R}^{2}\varepsilon_\cPqb\varepsilon_{\Pq*}+\mathcal{R}(1-\mathcal{R})(\varepsilon_\cPqb+\varepsilon_\Pq)\varepsilon_{\Pq*}+
 (1-\mathcal{R})^{2}\varepsilon_\Pq\varepsilon_{\Pq*},
\label{eq:prob21}
\end{equation}
where $\varepsilon_{\Pq*}$ is the effective misidentification rate,
and is computed by taking into account the expected flavour composition
of the ``extra'' jets in the events (\ie those not matched to a top-quark decay).
The effective misidentification rate is derived specifically for each event category.
From simulation, these extra jets are expected to come mostly from light-flavour jets ($\approx$87\%).
For completeness, for the case in which no jet from top-quark decay is
reconstructed, the probability of observing two $\cPqb$-tagged jets is
\begin{equation}
 P_\text{2j,2t,0d}=\varepsilon_{\Pq*}^{2}.
\label{eq:prob20}
\end{equation}

For each dilepton channel and jet multiplicity,
analogous expressions are derived and combined using the
probabilities $\alpha_i$ of having $i$ reconstructed jets from top-quark decays.
Additional terms are added to extend the model to events with more than two jets.
All efficiencies are determined per event category, after convolving the
corrections from dijet events in the data with the
expected efficiencies ($\varepsilon_\Pq$ and $\varepsilon_\cPqb$)
and the simulated jet \pt spectrum.

For the measurement of $\mathcal{R}$, a binned-likelihood function is constructed using the model described above
and the observed $\cPqb$-tagging multiplicity in events with two,
three, or four observed jets in the different dilepton channels.
A total of 36 event categories, corresponding to different permutations of three lepton-flavour pairs, three jet multiplicities,
and up to four observed $\cPqb$-tagged jets are used (see Fig.~\ref{fig:btagmult}). The likelihood is generically written as
\ifthenelse{\boolean{cms@external}}{
\begin{multline}
\mathcal{L}(\mathcal{R}, f_{\ttbar}, k_\mathrm{st}, f_\text{correct},
\varepsilon_b, \varepsilon_q,\varepsilon_{\Pq*},\theta_i) =\\
\prod_{\ell\ell} \prod_{N_\text{jets}=2...4}\prod_{k=0}^{N_\text{jets}}
{\mathcal P} [N_\text{ev}^{\ell\ell, N_\text{jets}}(k),
\hat{N}_\text{ev}^{\ell\ell,N_\text{jets}}(k)]\times\\
\prod_{i} \mathcal{G} (\theta_i^0,\theta_i,1),
\label{eq:hfcll}
\end{multline}
}{
\begin{equation}
\mathcal{L}(\mathcal{R}, f_{\ttbar}, k_\mathrm{st}, f_\text{correct},
\varepsilon_b, \varepsilon_q,\varepsilon_{\Pq*},\theta_i) =
\prod_{\ell\ell} \prod_{N_\text{jets}=2...4}\prod_{k=0}^{N_\text{jets}}
{\mathcal P} [N_\text{ev}^{\ell\ell, N_\text{jets}}(k),
\hat{N}_\text{ev}^{\ell\ell,N_\text{jets}}(k)]
\prod_{i} \mathcal{G} (\theta_i^0,\theta_i,1),
\label{eq:hfcll}
\end{equation}
}
where $N_\text{ev}^{\ell\ell, N_\text{jets}}$ ($\hat{N}_\text{ev}^{\ell\ell, N_\text{jets}}$) is the number of observed (expected)
events with $k$~$\cPqb$-tagged jets in a given dilepton channel
($\ell\ell={\Pe\Pe},\mu\mu,\Pe\mu$) with a given jet multiplicity ($N_\text{jets}$),
$\theta_i$ are the nuisance parameters (a total of 33, which will be discussed later),
and
${\mathcal G}$ is a Gaussian distribution. For the nominal fit, the nuisance
parameters are assumed to be unbiased ($\theta_i^0=0$) and normally distributed.
The nuisance parameters parameterise the effect of uncertainties, such as JES and JER, $\cPqb$-tagging
and misidentification rates, and $\mu_\mathrm{R}/\mu_\mathrm{F}$ scales,
amongst others, on the input parameters of the likelihood function.
The most likely value for $\mathcal{R}$ is found after profiling the
likelihood using the same technique described in Section~\ref{sec:samplecomposition}.
The result of the fit is verified to be unbiased in simulation,
by performing pseudo-experiments with dedicated MC samples where
$\mathcal{R}$ is varied in the [0,1] interval.
The residual difference found from these tests is assigned as a model calibration uncertainty.

\subsection{Measurement of \texorpdfstring{$\mathcal{R}$}{R}}
\label{sec:rmeasurement}

\begin{figure}[!hbtp]
\begin{center}
\includegraphics[width=\cmsFigWidth]{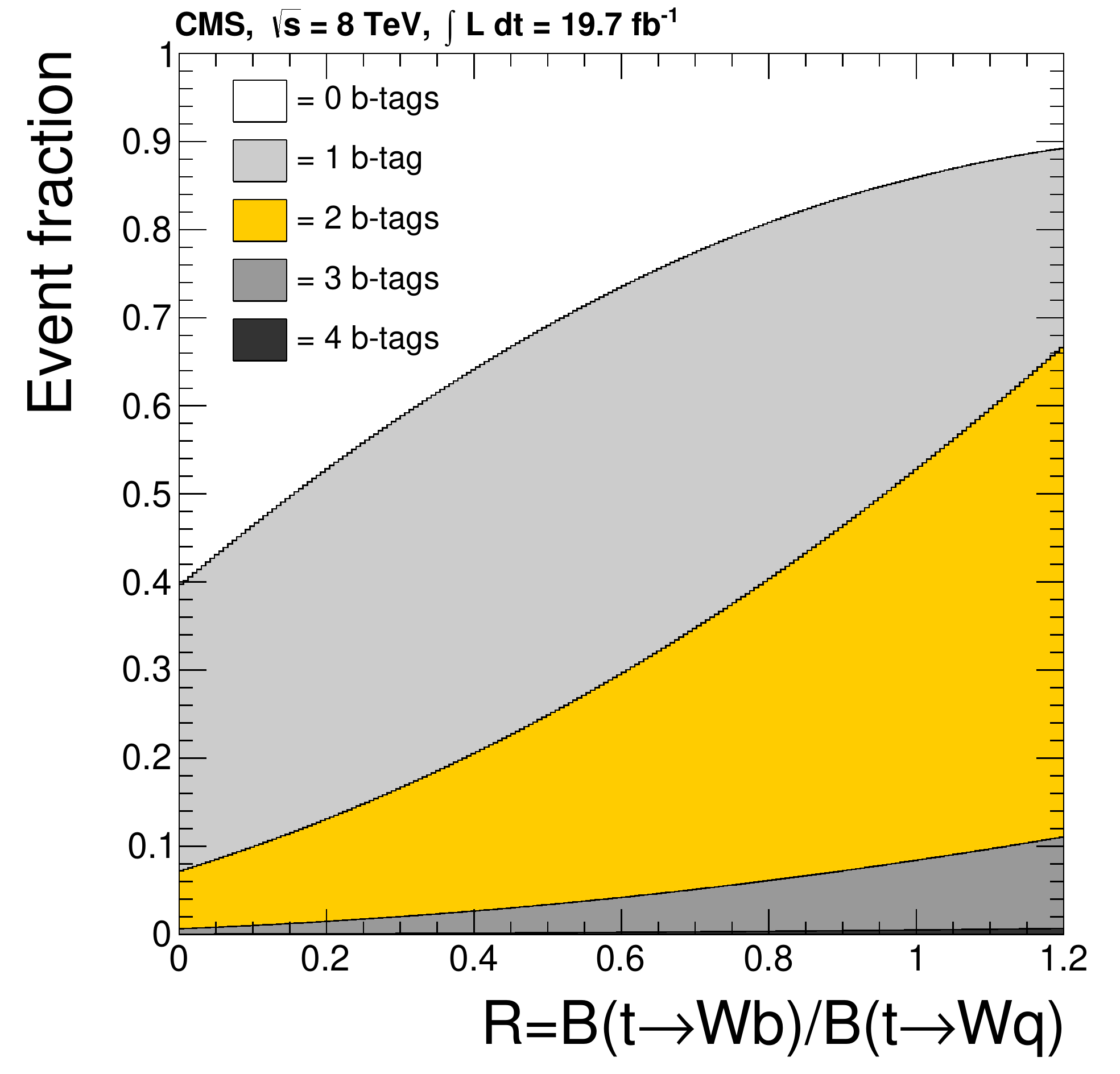}
\caption{
Expected event fractions of different $\cPqb$-tagged jet multiplicities in dilepton events as a function of $\mathcal{R}$.
}
\label{fig:rmodel}
\end{center}
\end{figure}

In the fit, $\mathcal{R}$ is allowed to vary without constraints. The
parameters of the model are all taken from the data:
$f_{\ttbar}$ and $k_\mathrm{st}$ are taken from Table~\ref{tab:samplepurity},
$f_\text{correct}$ is taken from Table~\ref{tab:correctcontributions},
$\varepsilon_\cPqb$ and $\varepsilon_\Pq$ from dijet-based
measurements~\cite{Chatrchyan:2012jua},
and $\varepsilon_{\Pq*}$ is derived following the method described in the
previous section.
Figure~\ref{fig:rmodel} shows the resulting prediction for the
fraction of events with different numbers of observed
$\cPqb$-tagged jets as a
function of $\mathcal{R}$. The individual predictions for all categories are summed to build the
inclusive model for the observation of up to four $\cPqb$-tagged jets in the selected events.

\begin{figure}[!hbtp]
\centering
\includegraphics[width=\cmsFigWidth]{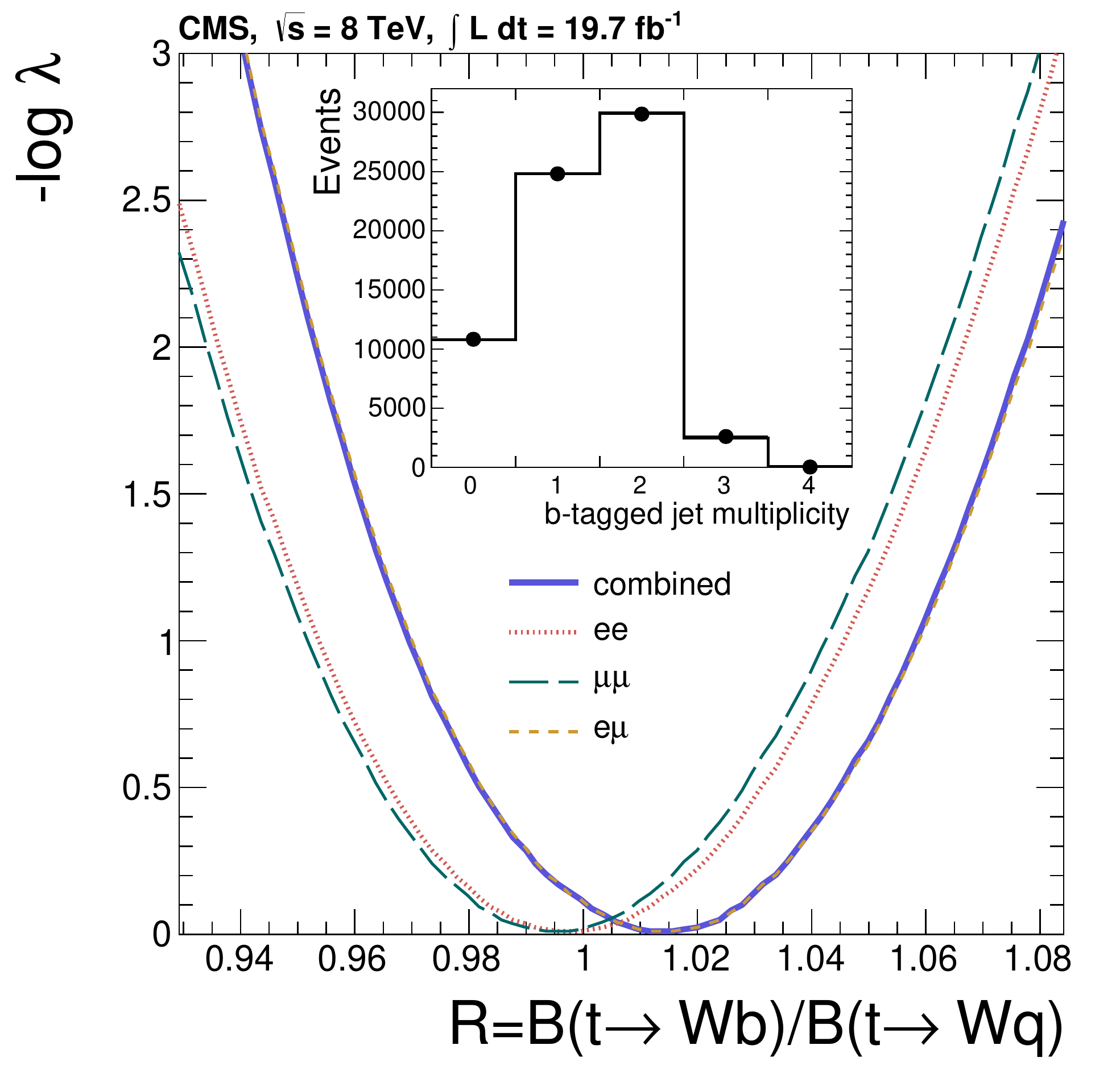}
\caption{
Variation of the log of the profile likelihood ratio ($\lambda$)
used to extract $\mathcal{R}$ from the data.
The variations observed in the combined fit
and in the exclusive ee, $\mu\mu$, and $\Pe\Pgm$ channels,
are shown.
The inset shows the inclusive $\cPqb$-tagged jet multiplicity
distribution and the fit distribution.
}
\label{fig:rfit}
\end{figure}

Figure~\ref{fig:rfit} shows the results obtained by maximising the profile likelihood.
The combined measurement of $\mathcal{R}$ gives
$\mathcal{R}=1.014\pm0.003\stat\pm0.032\syst$,
in good agreement with the SM prediction.
Fits to the individual channels give consistent results.
For these, we obtain values of
$\mathcal{R}_{\Pe\Pe}=0.997\pm 0.007\stat\pm 0.035\syst$,
$\mathcal{R}_{\mu\mu}=0.996\pm 0.007\stat\pm0.034\syst$, and
$\mathcal{R}_{\Pe\mu}=1.015\pm 0.003\stat\pm0.031\syst$
for the ee, $\mu\mu$, and $\Pe\Pgm$ channels, respectively.
The measurement in the $\Pe\Pgm$ channel dominates in the final combination
since the main systematic uncertainties are fully correlated and this
channel has the lowest statistical uncertainty.

The total relative uncertainty in the measurement of $\mathcal{R}$ is 3.2\%,
and is dominated by the systematic uncertainty,
whose individual contributions are summarized in
Table~\ref{tab:rfitsysts}.
The largest contribution to the systematic uncertainty is from
the $\cPqb$-tagging efficiency measurement.
Additional sources of uncertainty are related to the determination of the purity of the sample ($f_{\ttbar}$)
and the fraction of correct assignments ($f_\text{correct}$) from the data;
these quantities are affected by theoretical uncertainties related to the description of \ttbar events, which  have similar
impact on the final measurement, such as $\mu_\mathrm{R}/\mu_\mathrm{F}$, ME-PS, signal generator, top-quark mass, and top-quark \pt.
Instrumental contributions from JES and JER,
modelling of the unclustered \MET component in
simulation, and the contribution from the DY and
misidentified-lepton
backgrounds are each estimated to contribute a relative systematic
uncertainty $<0.6\%$.
Another source of uncertainty is due to the contribution from extra sources of heavy-flavour production,
either from gluon splitting in radiated jets or from decays in background events such as $\PW\to\cPqc\cPaqs$.
This effect has been estimated in the computation of $\varepsilon_{\Pq*}$ by assigning a conservative
uncertainty of 100\% to the $\cPqc$ and $\cPqb$ contributions. The effect of the uncertainty in the misidentification
efficiency is estimated to be small ($<1\%$), as well as other sources
of uncertainty, such as pileup and integrated luminosity.
After the fit is performed no nuisance parameter is observed to
change by more than $1.5\sigma$. The most relevant systematic
uncertainty ($\varepsilon_b$) is moved by $\sim 0.5\sigma$
as a result of the fit.

\begin{table}[htb]
\topcaption{Summary of the systematic uncertainties affecting the measurement of $\mathcal{R}$.
The values of the uncertainties are relative to the value of $\mathcal{R}$ obtained from the fit.
}
\label{tab:rfitsysts}
\centering
\resizebox{\cmsTabWidth}{!}{
\begin{tabular}{lc}\hline
Source & \multicolumn{1}{r}{Uncertainty (\%)}\\\hline
\multicolumn{2}{l}{Experimental uncertainties:} \\
$\varepsilon_\cPqb$ & 2.4 \\  
$\varepsilon_\Pq$	& 0.4 \\ 
$f_{\ttbar}$&	0.1 \\ 
DY &	0.2 \\ 
misidentified lepton	& 0.1 \\ 
JER &	0.5 \\ 
JES &	0.5 \\ 
unclustered \MET & 0.5 \\ 
integrated luminosity &	0.2 \\ 
pileup &	0.5 \\ 
simulation statistics &	0.5 \\ 
$f_\text{correct}$ &	0.5 \\ 
model calibration & 0.2 \\ 
selection efficiency & 0.1 \\\hline 
\multicolumn{2}{l}{Theoretical uncertainties:} \\
top-quark mass & 0.9 \\ 
top-quark \pt &	0.5  \\ 
ME-PS &	0.5  \\ 
$\mu_\mathrm{R}/\mu_\mathrm{F}$ & 0.5 \\ 
signal generator &	0.5 \\ 
underlying event &	0.1 \\ 
colour reconnection & 0.1 \\ 
hadronisation &	0.5 \\ 
PDF	& 0.1 \\ 
$\cPqt\to\PW\Pq$ flavour  & 0.4 \\ 
$\abs{V_{\cPqt\cPqd}} / \abs{V_{\cPqt\cPqs}}$ & $<$0.01 \\
relative single-top-quark fraction (tW) &	0.1 \\
VV (theoretical cross section) & 0.1 \\ 
extra sources of heavy flavour &	0.4 \\\hline 
{Total systematic} &	3.2 \\\hline 
\end{tabular}
}
\end{table}

\begin{figure}[!hbtp]
\centering
\includegraphics[width=\cmsFigWidth]{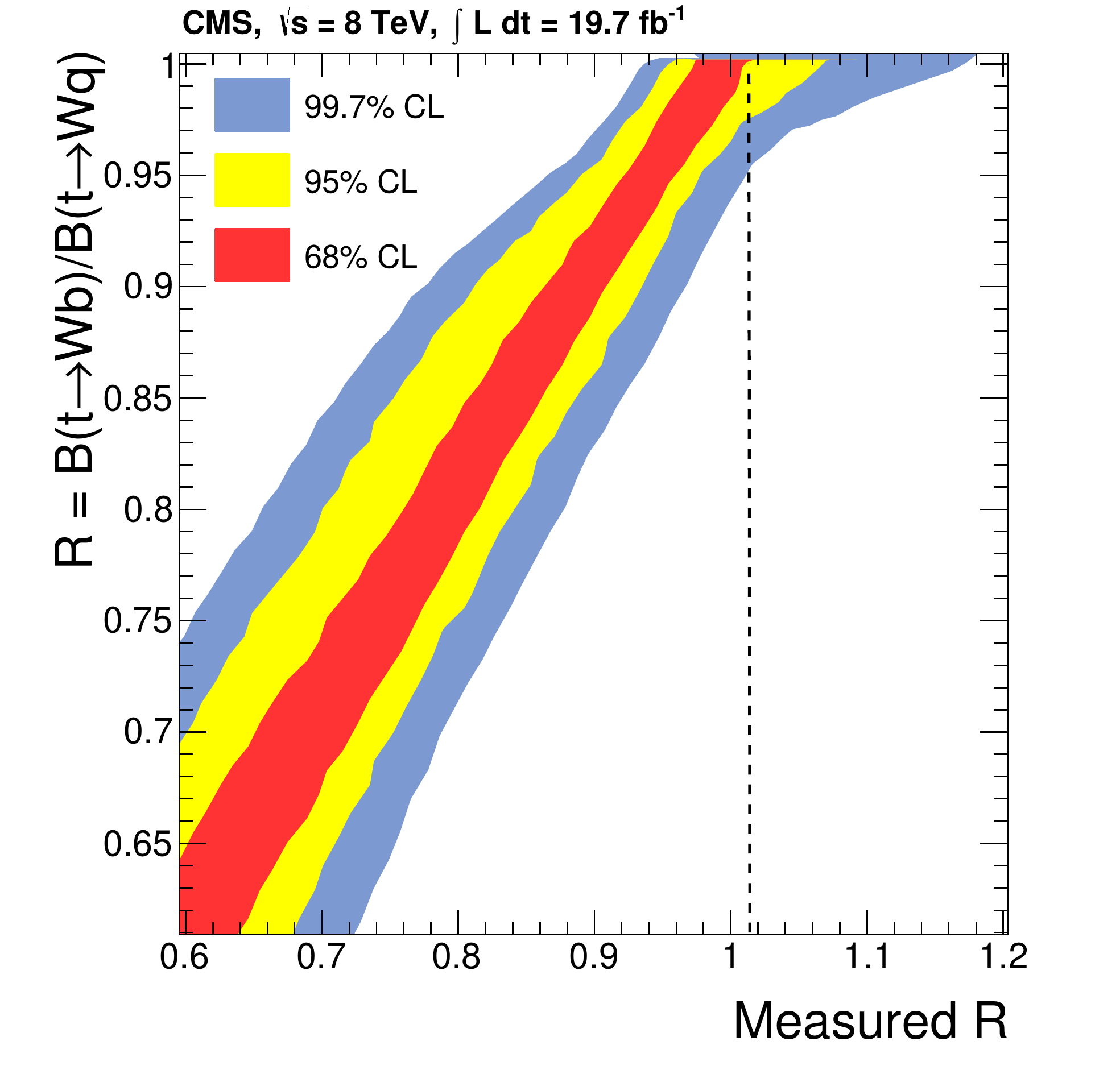}
\caption{Expected limit bands at different confidence levels
as a function of the measured $\mathcal{R}$ value.
The range of measured values of $\mathcal{R}$  that are
allowed for each true value of $\mathcal{R}$
are shown as coloured bands for different confidence levels.
The observed value of $\mathcal{R}$ is shown as the dashed line.
}
\label{fig:rlimits}
\end{figure}

If the three-generation CKM matrix is assumed to be unitary, then $\mathcal{R}=\abs{V_{\cPqt\cPqb}}^2$~\cite{Alwall:2006bx}.
By performing the fit in terms of $\abs{V_{\cPqt\cPqb}}$, a value of
$\abs{V_{\cPqt\cPqb}}=1.007\pm0.016\,\text{(stat.+syst.)}$ is measured.
Upper and lower endpoints of the 95\% CL interval for $\mathcal{R}$ are extracted by
using the Feldman--Cousins (FC) frequentist approach~\cite{Feldman:1997qc}.
The implementation of the FC method in \textsc{RooStats}~\cite{Moneta:2010pm}
is used to compute the interval. All the nuisance parameters (including $\varepsilon_\cPqb$) are profiled
in order to take into account the corresponding uncertainties (statistical and systematic).
If the condition $\mathcal{R}\leq 1$ is imposed, we obtain
$\mathcal{R}>0.955$ at the 95\% CL.
Figure~\ref{fig:rlimits} summarizes the expected limit bands for 68\% CL, 95\% CL, and
99.7\% CL, obtained from the FC method.
The expected limit bands are determined from the distribution of the profile
likelihood obtained from simulated pseudo-experiments.
The upper and lower acceptance regions constructed in this procedure are used
to determine the endpoints on the allowed interval for $\mathcal{R}$.
In the pseudo-experiments the expected signal and background yields
are varied using Poisson probability distributions for the statistical
uncertainties and Gaussian distributions for the systematic uncertainties.
By constraining $\abs{V_{\cPqt\cPqb}}\leq 1$,
a similar procedure is used to obtain $\abs{V_{\cPqt\cPqb}}>0.975$ at the 95\% CL.

\subsection{Indirect measurement of the top-quark total decay width}
\label{sec:topwidthmeasurement}

The result obtained for $\mathcal{R}$ can be combined with a
measurement of the single-top-quark production cross section in the
$t$-channel to yield an indirect determination of the top-quark
total width $\Gamma_{\cPqt}$.
Assuming that $\sum_\Pq \mathcal{B}(\cPqt\to\PW\Pq)=1$, then
$\mathcal{R}=\mathcal{B}(\cPqt\to\PW\cPqb)$ and
\begin{equation}
\Gamma_{\cPqt} = \frac{\sigma_{t\text{-ch.}}}{\mathcal{B}(\cPqt\to\PW\cPqb)} \cdot \frac{\Gamma(\cPqt\to\PW\cPqb)}{\sigma_{t\text{-ch.}}^\text{theor.}},
\label{eq:gammatot}
\end{equation}
where $\sigma_{t\text{-ch.}}$ ($\sigma_{t\text{-ch.}}^\text{theor.}$) is the measured (theoretical) $t$-channel
single-top-quark cross section
and $\Gamma(\cPqt\to\PW\cPqb)$ is the top-quark partial decay
width to $\PW\cPqb$.
If we assume a top-quark mass of 172.5\GeV,
then the theoretical partial width of the top quark decaying to Wb is
$\Gamma(\cPqt\to\PW\cPqb)=1.329$\GeV~\cite{Beringer:1900zz}.
A fit to the $\cPqb$-tagged jet multiplicity distribution in
the data
is performed, leaving $\Gamma_{\cPqt}$ as a free parameter.
In the likelihood function we use the theoretical prediction for the
$t$-channel cross section at $\sqrt{s}=7\TeV$
from Ref.~\cite{Kidonakis:2011wy} and the corresponding CMS measurement
from Ref.~\cite{Chatrchyan:2012ep}.
The uncertainties in the predicted and measured cross sections are taken into account as extra nuisance
parameters in the fit. The uncertainty in the theoretical cross section is parameterised by convolving a
Gaussian function for the PDF uncertainty with a uniform prior describing the factorisation and renormalisation
scale uncertainties. Some uncertainties in the experimental
cross section measurement such as those from JES and JER,
$\cPqb$-tagging efficiency, $\mu_\mathrm{R}/\mu_\mathrm{F}$ scales, and
ME-PS threshold for \ttbar production are
fully correlated with the ones assigned to the measurement of
$\mathcal{R}$.
All others are summed in quadrature and assumed to be uncorrelated.
After performing the maximum-likelihood fit, we measure
$\Gamma_{\cPqt}=1.36\pm0.02\stat^{+0.14}_{-0.11}\syst\GeV$,
in good agreement with the theoretical expectation~\cite{Beringer:1900zz}.
The dominant uncertainty comes from the measurement of the $t$-channel cross section,
as summarized in  Table~\ref{tab:gammatsysts}.

\begin{table}[htp]
\topcaption{Summary of the systematic uncertainties in the
  measurement of $\Gamma_{\cPqt}$.
The values of the uncertainties are relative to the value of
$\Gamma_{\cPqt}$ obtained from the fit.
The ``Other sources'' category combines all the individual contributions below 0.5\%.
}
\label{tab:gammatsysts}
\centering
\resizebox{\cmsTabWidth}{!}{
\begin{tabular}{lc}\hline
Source & \multicolumn{1}{r}{Uncertainty (\%)}\\\hline
Single-top quark $t$-channel cross section & 9.2 \\
$\varepsilon_\cPqb$ & 4.3 \\
JES &	0.7 \\
pileup & 0.8\\
ME-PS & 0.8\\
$\mu_\mathrm{R}/\mu_\mathrm{F}$ & 0.8\\
top-quark mass & 0.6 \\
Other sources & 1.5 \\\hline
{Total systematic} & 10.4 \\\hline
\end{tabular}
}
\end{table}

\section{Summary}
\label{sec:conclusions}

A measurement of the ratio of the top-quark branching fractions
$\mathcal{R}=\mathcal{B}(\cPqt\to\PW\cPqb)/\mathcal{B}(\cPqt\to\PW\Pq)$,
where the denominator includes the sum over the
branching fractions of the top quark to a $\PW$ boson and a
down-type quark (q = b, s, d),
has been performed using a sample of \ttbar dilepton events.
The sample has been selected from proton-proton collision data
at $\sqrt{s}=8\TeV$ from an integrated luminosity of 19.7\fbinv,
collected with the CMS detector.
The $\cPqb$-tagging and misidentification efficiencies are derived from multijet control samples.
The fractions of events with 0, 1, or 2 selected jets from top-quark decays are determined
using the lepton-jet invariant-mass spectrum and an empirical
model for the misassignment contribution.
The unconstrained measured value of
$\mathcal{R}=1.014\pm0.003\stat\pm0.032\syst$
is consistent with the SM prediction, and the main systematic uncertainty is
from the $\cPqb$-tagging efficiency ($\approx$2.4\%).
All other uncertainties are $<1\%$.
A lower limit of $\mathcal{R}> 0.955$ at 95\% CL is obtained after
requiring $\mathcal{R}\leq 1$ and taking into account both statistical and systematical uncertainties.
This result translates into a lower limit $\abs{V_{\cPqt\cPqb}} > 0.975$ at 95\% CL when assuming the unitarity of the three-generation CKM matrix.
By combining this result with a previous CMS measurement of the $t$-channel
production cross section for single top quarks, an indirect
measurement of the top-quark total decay width
$\Gamma_{\cPqt}=1.36\pm0.02\stat^{+0.14}_{-0.11}\syst\GeV$
is obtained, in agreement with the SM expectation.
These measurements of $\mathcal{R}$ and $\Gamma_{\cPqt}$
are the most precise to date and the first obtained at the LHC.

\section*{Acknowledgements}

We congratulate our colleagues in the CERN accelerator departments for the excellent performance of the LHC and thank the technical and administrative staffs at CERN and at other CMS institutes for their contributions to the success of the CMS effort. In addition, we gratefully acknowledge the computing centres and personnel of the Worldwide LHC Computing Grid for delivering so effectively the computing infrastructure essential to our analyses. Finally, we acknowledge the enduring support for the construction and operation of the LHC and the CMS detector provided by the following funding agencies: BMWFW and FWF (Austria); FNRS and FWO (Belgium); CNPq, CAPES, FAPERJ, and FAPESP (Brazil); MES (Bulgaria); CERN; CAS, MoST, and NSFC (China); COLCIENCIAS (Colombia); MSES and CSF (Croatia); RPF (Cyprus); MoER, SF0690030s09 and ERDF (Estonia); Academy of Finland, MEC, and HIP (Finland); CEA and CNRS/IN2P3 (France); BMBF, DFG, and HGF (Germany); GSRT (Greece); OTKA and NIH (Hungary); DAE and DST (India); IPM (Iran); SFI (Ireland); INFN (Italy); NRF and WCU (Republic of Korea); LAS (Lithuania); MOE and UM (Malaysia); CINVESTAV, CONACYT, SEP, and UASLP-FAI (Mexico); MBIE (New Zealand); PAEC (Pakistan); MSHE and NSC (Poland); FCT (Portugal); JINR (Dubna); MON, RosAtom, RAS and RFBR (Russia); MESTD (Serbia); SEIDI and CPAN (Spain); Swiss Funding Agencies (Switzerland); MST (Taipei); ThEPCenter, IPST, STAR and NSTDA (Thailand); TUBITAK and TAEK (Turkey); NASU and SFFR (Ukraine); STFC (United Kingdom); DOE and NSF (USA).

Individuals have received support from the Marie-Curie programme and the European Research Council and EPLANET (European Union); the Leventis Foundation; the A. P. Sloan Foundation; the Alexander von Humboldt Foundation; the Belgian Federal Science Policy Office; the Fonds pour la Formation \`a la Recherche dans l'Industrie et dans l'Agriculture (FRIA-Belgium); the Agentschap voor Innovatie door Wetenschap en Technologie (IWT-Belgium); the Ministry of Education, Youth and Sports (MEYS) of Czech Republic; the Council of Science and Industrial Research, India; the Compagnia di San Paolo (Torino); the HOMING PLUS programme of Foundation for Polish Science, cofinanced by EU, Regional Development Fund; and the Thalis and Aristeia programmes cofinanced by EU-ESF and the Greek NSRF.

\bibliography{auto_generated}   
\cleardoublepage \appendix\section{The CMS Collaboration \label{app:collab}}\begin{sloppypar}\hyphenpenalty=5000\widowpenalty=500\clubpenalty=5000\textbf{Yerevan Physics Institute,  Yerevan,  Armenia}\\*[0pt]
V.~Khachatryan, A.M.~Sirunyan, A.~Tumasyan
\vskip\cmsinstskip
\textbf{Institut f\"{u}r Hochenergiephysik der OeAW,  Wien,  Austria}\\*[0pt]
W.~Adam, T.~Bergauer, M.~Dragicevic, J.~Er\"{o}, C.~Fabjan\cmsAuthorMark{1}, M.~Friedl, R.~Fr\"{u}hwirth\cmsAuthorMark{1}, V.M.~Ghete, C.~Hartl, N.~H\"{o}rmann, J.~Hrubec, M.~Jeitler\cmsAuthorMark{1}, W.~Kiesenhofer, V.~Kn\"{u}nz, M.~Krammer\cmsAuthorMark{1}, I.~Kr\"{a}tschmer, D.~Liko, I.~Mikulec, D.~Rabady\cmsAuthorMark{2}, B.~Rahbaran, H.~Rohringer, R.~Sch\"{o}fbeck, J.~Strauss, A.~Taurok, W.~Treberer-Treberspurg, W.~Waltenberger, C.-E.~Wulz\cmsAuthorMark{1}
% ...
\vskip\cmsinstskip
% ...
1:~~Also at Vienna University of Technology, Vienna, Austria\\
2:~~Also at CERN, European Organization for Nuclear Research, Geneva, Switzerland\\
% ...

\end{sloppypar}
\end{document}